\documentclass[emulateapj,twocolumn]{aastex63}
\usepackage{amsmath}
\usepackage{graphicx}
\usepackage{natbib}
\usepackage[T1]{fontenc}
\usepackage{booktabs}
\usepackage{lipsum}

\usepackage{txfonts} %use times font for math
\usepackage[figure,figure*]{hypcap} %Figure refs go figures
\usepackage{chngcntr}

\begin{document}
\title{An Improved and Physically-Motivated Scheme for Matching Galaxies with Dark Matter Halos}
\author{Stephanie Tonnesen}
\affil{Flatiron Institute, CCA, 162 5th Avenue, New York, NY 10010 USA}
\email{stonnes@gmail.com (ST)}

\author{Jeremiah P. Ostriker}
\affil{Flatiron Institute, CCA, 162 5th Avenue, New York, NY 10010 USA}
\affil{Princeton University Observatory, Ivy Lane, Princeton, NJ 08544 USA}
\affil{Columbia Astrophysics Laboratory, 538 W 120th St, New York, NY 10027 USA}

\date{January 2021}

\begin{abstract}
The simplest scheme for predicting real galaxy properties after performing a dark matter simulation is to rank order the real systems by stellar mass and the simulated systems by halo mass and then simply assume monotonicity  - that the more massive halos host the more massive galaxies. This has had some success, but we study here if a better motivated and more accurate matching scheme is easily constructed by looking carefully at how well one could predict the simulated IllustrisTNG galaxy sample from its dark matter computations. We find that using the dark matter rotation curve peak velocity, $v_{max}$, for normal galaxies reduces the error of the prediction by 30\% (18\% for central galaxies and 60\% for satellite systems) - following expectations from the physics of monolithic collapse. For massive systems with halo mass $>$ 10$^{12.5}$ M$_{\odot}$ hierarchical merger driven formation is the better model and dark matter halo mass remains the best single metric. Using a new single variable that combines these effects, $\phi$ $=$ $v_{max}$/$v_{max,12.7}$ + M$_{peak}$/(10$^{12.7}$ M$_{\odot}$) allows further improvement and reduces the error, as compared to ranking by dark matter mass at $z=0$ by another 6\% from $v_{max}$ ranking.  Two parameter fits -- including environmental effects produce only minimal further impact. 

\end{abstract}

\section{Introduction}

There has been a great deal of progress in recent years in our understanding and simulations of galaxy formation. The initial conditions seem to be well specified by LCDM cosmological models and their variants, and hydrodynamic codes now have the capacity and resolution to include many of the physical processes relevant to galaxy formation and evolution. These include dark matter gravitational collapse, gas cooling, star formation and the mechanical and radiative feedback from stars and central, massive black holes (BHs). There is no widely accepted mechanism for the formation of these massive BHs, but, assuming their formation as seed BHs, their evolution and effects on the surrounding galaxies are now reasonably well modeled. Recent summaries of the status of existing work on galaxy formation and evolution are presented in \citet{Somerville2015} and \citet{Naab2017}. For more detailed treatments one can consult the EAGLE \citep{Schaye2015,Crain2015}, FIRE \citep{Hopkins2018a,Hopkins2018b}, IllustrisTNG \citep{Vogelsberger2014,Genel2014,Weinberger2017,Pillepich2018}, MUFASA \citep{Dave2016}, NIHAO \citep{Wang2015,Blank2019} and other simulations. 

But there are virtues to considering simpler treatments that do not need to rely on sub-grid modeling and which are easily adaptable to analyzing large data sets. Of these, the simplest, perhaps, is the abundance matching scheme which is based on the fact that, in all variants of the LCDM modeling, galaxies live in more massive dark matter (DM) halos, quasi-spherical lumps of dark matter, which grow via gravitational instabilities from very low amplitude ( 10$^{-5}$), gaussian perturbations imparted at very early times in a roughly power law distribution by unknown processes (thought to be related to inflation) ((\citealp[e.g.][]{Vale2004, Vale2006, Vale2008}).

There is good agreement on how to compute the formation of these dark matter halos with several computational codes now able to make moderately high-resolution cosmic scale volumes containing accurate distributions of DM halos having well defined properties. Early analyses by
\citet{Navarro1997} showed that these could be represented to a reasonable approximation by three numbers, a mass, a virial radius and a core radius, with the ratio of the latter two numbers represented as the concentration.  To zeroth order observed galaxies can be represented by their stellar masses, or alternatively by their stellar luminosities.  

Thus, the simplest possible scheme for populating a volume in the universe with galaxies would be to populate it first with DM halos, and then make a rank ordered list of these with the most massive first. Next, one could take the same volume from the real universe and rank order the observed galaxies by mass (or luminosity), and then simply assume that the more massive halos hosted more massive galaxies, putting in each halo the corresponding galaxy.

This, almost ridiculously simple, scheme was pursued by Vale \& Ostriker in three papers (\citealp{Vale2004, Vale2006, Vale2008}; \citealp[see also][]{Kravtsov2004, Tasitsiomi2004}).
Using this scheme (or a variation thereof), one can take different variants of the LCDM model, compute the halo distribution, populate the halos with galaxies using the simple abundance matching scheme and then compare to observations. By construction, the luminosity functions must come out to be correct, but correlation functions \citep{Conroy2006,Marin2008,Guo2010,TrujilloGomez2011}, pair counts \citep{Berrier2006}, magnitude gap statistics \citep{Hearin2013a, Ostriker2019}, and galaxy-galaxy lensing \citep{Hearin2013b} can be usefully compared to observations.  

Two immediate questions arise.  First, is there a better zeroth order scheme than ranking by mass and matching.  Several authors have considered this question.  For example, while the DM in subhalos is quickly stripped once they are accreted onto halos, stripping of the more centralized galaxy starts later \citep{Nagai2005}, and in fact their optical sizes are observed to grow with cosmic time (\citealp[cf][]{vanDokkum2010}), presumably due to the accretion of smaller satellite systems.  Thus, \citet{Conroy2006} improved subhalo abundance matching by matching galaxies to halos at the time at which they are accreted onto a central halo.  Even earlier, \citet{Kravtsov2004} proposed using the maximum circular velocity of (sub)halos, $v_{max}$, which is more stable than halo mass to stripping.  \citet{Reddick2013} tested abundance matching models using several halo properties, and found that only v$_{peak}$ (defined as the peak value of $v_{max}$ over the history of the halo) or a combination of $v_{max}$ for central galaxies and v$_{peak}$ for satellite galaxies is able to reproduce observations of galaxy clustering.  Indeed, \citet{Zentner2014} argues that abundance matching using $v_{max}$ matches several observed galaxy statistics \citep{Conroy2006,Hearin2013a,Hearin2013b, Reddick2013} because halo mass alone does not determine the halo velocity profile.

\citet{Xu2018} confirmed that for the central galaxies in the original Illustris simulations, M$_*$ is more tightly correlated with v$_{peak}$ than with halo mass.  They also find that at fixed v$_{peak}$, the correlation between M$_*$ and other halo properties is removed. %(halo mass, concentration, half mass formation time, and the halo mass (specific) accretion rate near z=0).  
In \citet{He2020}, the author uses subhalo abundance matching and finds that  v$_{peak}$ correlates best with the stellar mass at the epoch of v$_{peak}$ in both central and satellite galaxies in EAGLE, Illustris, and IllustrisTNG.  Stellar mass stripping of satellite galaxies results in increased scatter in the $z=0$ M$_{*}$ to v$_{peak}$ relation.  \citet{ChavesMontero2016} find that $v_{relax}$, defined as the maximum of the circular
velocity of a dark matter structure while it fulfils a relaxation criterion, as evaluated along its entire history, correlates most strongly with M$_*$.  

Second, does there exist any first order refinements of the mass-matching scheme that could be implemented, which would be easy to apply and would significantly increase its accuracy. In fact, \citet{Lehmann2017} point out that while ranking by $v_{max}$ is similar to ranking by halo mass, at fixed halo mass more concentrated halos have higher $v_{max}$ \citep{Klypin2011}.  They therefore use a parameterization from \citet{Mao2015} that includes both halo mass and concentration: 
\begin{equation}
V_{\alpha} \equiv V_{vir}(\frac{V_{max}}{V_{vir}})^{\alpha}
\end{equation}
where $v_{max}$ is the maximal circular velocity of the halo and 
\begin{equation}
V_{vir} \equiv (\frac{GM_{vir}}{R_{vir}})^{1/2}
\end{equation}

They find that an $\alpha$ $\sim$ 0.57, with a scatter of 0.17 dex is the best fit to the SDSS clustering measurements, indicating that both the halo mass and the maximum circular velocity impact the stellar mass of galaxies.  

Identifying the variables causing the scatter in an abundance matching model also leads to a better understanding of the important physical processes influencing galaxy formation.  For example, \citet{Matthee2017} matched galaxies in the hydrodynamic EAGLE simulation to those in the dark matter only simulations, and found that much of the scatter is due to halo concentration.  Specifically, at a constant halo mass, higher concentration is correlated to higher stellar mass, likely because higher concentrations imply earlier formation times.  However, they do not find a halo property than can explain the remaining scatter.  \citet{Martizzi2020} considered the influence of formation time and environment on the scatter in the stellar mass to halo mass relation using IllustrisTNG.  Sorting by the current subhalo mass, they find that the scatter in the relation for central galaxies is correlated more strongly with formation time, while the scatter for satellite galaxies is correlated more strongly with environment.  On the other hand, \citet{Dragomir2018} compare the galaxy luminosity functions and the galaxy stellar mass function at different environmental densities for SDSS observations and a subhalo abundance matching model applied to the Bolshoi-Planck simulation and find that the model predictions agree well with observations.

In this paper we will try to both physically motivate and improve the simplest matching scheme.  We will use IllustrisTNG to determine whether complicating the most simple form of subhalo abundance matching, ie matching stellar mass to a single halo property, reduces scatter in the assignment of galaxies to halos.  We first verify the halo property that produces the least scatter in the relation, $v_{max}$, considering the central and satellite populations separately and selecting different mass ranges even in this initial step.  We then provide a physical motivation for the parameters used in the optimal matching scheme.  Then, similarly to \citet{Martizzi2020}, we calculate how the scatter is reduced when we fold in a second halo feature.  However, unlike these recent works, we test a wide range of possible parameters and possible combinations thereof.  

In Section \ref{sec:method} we describe our sample of galaxies and the variables we measure for each galaxy.  %methodology for testing and evaluating procedures for using matching techniques to predict galaxy properties given dark matter simulations. 
In Section \ref{sec:theory} we outline a theoretical basis for selecting the dark matter halo property on which to rank, and in section \ref{sec:ranking} we try out the simplest one parameter schemes and show that the physically motivated focus on peak velocity dispersion is best for normal galaxies but that total halo mass remains best for first brightest, massive, central systems as would be expected from the presented physical arguments. Section \ref{sec:secondary_variables} broadens the treatment to include multiple variables - including environment - and then in Sections \ref{sec:discussion} and \ref{sec:tests} we present an overall discussion of the results and possible tests of the scheme.  Finally, in Section \ref{sec:conclusions} we summarize our findings and conclusions.

\section{Methods}\label{sec:method}
\subsection{IllustrisTNG}
The IllustrisTNG100 \citep[public data release: ][]{ Nelson2019}\footnote{www.tng-project.org} is part of a suite of cosmological simulations run using the AREPO moving mesh code \citep{Springel2010}.  TNG100 has a volume of  110.7 Mpc$^3$ and a mass resolution of $7.5 \times 10^6 \mathrm{M}_{\odot}$ and $1.4 \times 10^6 \mathrm{M}_{\odot}$ for dark matter and baryons, respectively.  The TNG suite implements upgraded subgrid models compared to the Illustris simulation \citep{Vogelsberger2014, Genel2014}; specifically, a modified black hole accretion and feedback model \citep{Weinberger2017}, updated galactic winds \citep{Pillepich2018}.  TNG also includes magnetohydrodynamics \citep{Pakmor2011}. 
%The gravitational softening is $0.74$ kpc at $z=0$ for the collisionless particles and adaptive with a minimum of $0.18$ kpc at $z=0$ for the gas cells.  The gas cells have a minimum (median) radius of 14 pc (15.8 kpc), and star-forming gas cells have a mean radius of 355 pc.

\subsection{Galaxy Selection}\label{sec:galaxy_selection}

We use galaxy populations from the IllustrisTNG 100 simulation described above.  We consider galaxies at the $z=0$ output with dark matter masses of  $\ge$10$^{11}$ M$_{\odot}/h$ in the dark-matter only run (DMO) that are matched in the full hydrodynamical simulation with galaxies whose stellar mass is greater than 10$^{9}$ M$_{\odot}/h$.

In detail, we first selected all galaxies in the DMO simulation with a dark matter mass greater than 10$^{11}$ M$_{\odot}$/$h$.  We then used the publicly available matching data to find the corresponding galaxy in the full hydro simulation.  We use all galaxies identified with masses above 5 $\times$ 10$^6$ M$_{\odot}$/$h$, which clearly includes galaxies that are underresolved in the simulation.  However, at our minimum dark matter halo mass, the lowest stellar mass of any galaxy in our sample is 7 $\times$ 10$^7$ M$_{\odot}$/$h$.  In order to only include well-resolved galaxies, that are more likely to be in an observational sample, our final analysis only includes galaxies with stellar masses above 10$^9$ M$_{\odot}$/$h$.

\subsection{Galaxy Environmental Measures}

We use nearby galaxies to measure the local environment.  We include all galaxies with dark matter masses above 10$^9$ M$_{\odot}/h$ within 1 Mpc, 2 Mpc, 5 Mpc, 8 Mpc or 15 Mpc of each galaxy.  In order to have a more physical measure of the local mass density, we summed the total mass of all these galaxies.  

We also were able to separate galaxies into satellites or centrals using the GroupFirstSub identifier in the IllustrisTNG DMO simulation.  This allowed us to perform our fits for the entire sample and for satellites and centrals separately, and, as we shall see, the two categories are significantly different in their properties.  This gives us three samples: ``all", ``centrals" and ``satellites".  The fourth sample is labeled as ``mix", which is the combined sample in which satellites and centrals are fit separately.

\subsection{Concentration}\label{sec:concentration}

We use three measures of the concentration.  First, from \citet{Bose2019} we use:
\begin{equation}\label{eqn:concentrationvmax}
c_v \equiv \frac{V_{max}}{H_oR_{max}}
\end{equation}

\noindent where $v_{max}$ is the maximum velocity of the simulated rotation curve and R$_{max}$ is the radius at which V$_c$ is maximal.  
\citet{Bose2019} show that this is equivalent to the concentration calculated using all the particles in a halo and assuming an NFW profile (\citealp[see also][]{Moline2017}).  

Second, we use the ratio $c_h$$\equiv$$v_{max}$/V$_{halfmass}$, where V$_{halfmass}$ is the circular velocity at the half mass radius of the dark matter halo, calculated as $\sqrt{GM_{halfmass}/R_{halfmass}}$.

Finally, we use the ratio $c_R$$\equiv$R$_{max}$/R$_{halfmass}$.

\subsection{Percent Error}\label{sec:error}

We define the error as: 
\begin{equation}
Error \equiv \frac{\sum\limits_N \mid log(M_{true}/M_{prediction})\mid}{N}
\end{equation}

\noindent so that for small errors our definition is equivalent to 0.43 times the average fractional error.

\section{Rank Ordering}

In this section we discuss using rank ordering to match halos and galaxies.  Specifically, we first present a straightforward theoretical scaffold for selecting the halo property best suited for rank ordering.  Then, using IllustrisTNG we confirm our derivation.

\subsection{A Simple Theoretical Basis for Selecting the Ordering Halo Property}\label{sec:theory}

We can begin with an assumption that stellar mass is related to the baryonic mass scaled to the dark matter mass, corrected by the fraction of matter that cools and forms stars in the center of the halo:

\begin{equation}\label{eqn:theorystart}
M_* \propto M_{peak}\frac{\Omega_b}{\Omega_d} \frac{t_{form}}{t_{cool,form}}
\end{equation}

Here $t_{form}$ is the formation time of the halo and $t_{cool,form}$ is the time required for the baryons to cool and condense into a galaxy.  This is simply putting in the form of an equation the classical idea of ``monolithic collapse" first proposed by \citet{Eggen1962}.

We can relate the mean density of the galaxy to its mass and radius:

\begin{equation}
\rho_{max} \equiv \frac{M_{max}}{\frac{4}{3} \pi r_{max}^3}
\end{equation}

In which $\rho_{max}$, $M_{max}$, and $r_{max}$ are the density, mass, and radius at which the circular velocity reaches $v_{max}$, where $v_{max}^2$ $=$ $\frac{GM_{max}}{r_{max}}$.

We can relate t$_{form}$ to the halo density assuming standard gravitational collapse (Gunn \& Gott 1972):
\begin{equation}
G<\rho> \equiv t_{form}^{-2}
\end{equation}

We can also relate t$_{cool,form}$ to density using energy conservation and the standard cooling equations:

\begin{equation}
\frac{\frac{3}{2}kT_{max}}{m} \equiv \frac{GM_{max}}{r_{max}}
\end{equation}

The above equation defines T$_{max}$.  We subsequently can define t$_{cool,form}$ as:

\begin{equation}
\Lambda(T_{max})\rho_{max}^2 \equiv \frac{\frac{3}{2} \rho_{max} k T_{max}}{t_{cool,form}}
\end{equation}

Thus, t$_{cool,form}$ $\propto$ $\rho_{max}^{-1}f^{-1}$ where $f$$\equiv$$\Lambda(T_{max})/T_{max}$, and t$_{form}$ $\propto$ $\rho_{max}^{-\frac{1}{2}}$.  If we use these relations in Equation \ref{eqn:theorystart}, we find that 

\begin{equation}\label{eqn:mstarvmax4}
M_* \propto M_{peak}\rho_{max}^{\frac{1}{2}}f \propto (\frac{M_{max}}{r_{max}})^{\frac{3}{2}}f \propto v_{max}^3f
\end{equation}

For low mass halos for which $f$ is nearly proportional to T$_{max}$ this indicates a steep dependence on M$_*$ on $v_{max}$, which flattens for higher mass galaxies with higher T$_{max}$.

We highlight that this derivation is based on the assumption of spherical collapse of the halo and pure radiative cooling.  We do not consider any complicating processes that we know affect galaxies in the universe, such as mergers or feedback from star formation or AGN.  In fact, we might expect this relation between M$_*$ and $v_{max}^3f$ to break down more often for higher mass galaxies, as they have been found to have later growth times where these assumptions clearly breakdown \citep{Behroozi2013}.  For first brightest systems, sitting in massive halos from which they can accrete satellites, one would expect M$_{peak}$ to be more relevant, and, as we have noted, both observations (\citealp[e.g.][]{vanDokkum2010}) and LCDM theory argue that hierarchical accretion is the dominant process for first brightest galaxies.

Indeed, we can go a step farther and ask the basis for and the value of the transition mass above which ``normal" growth of the stellar component from a cooling collapse becomes difficult. This was addressed in a paper by  \citet{Rees1977}(eqn 20) in an elementary discussion of the maximum mass of cosmic gas that can cool and collapse in a dynamical time. They did not include the important effects of dark matter in their treatment and obtained a mass of $[(\frac{\hbar c}{Gm_p^2})^2 (\frac{e^2}{\hbar c})^5(\frac{m_p}{m_e})^{\frac{1}{2}}]m_p$$\sim$10$^{12}$  M$_{\odot}$ in baryons. Had the effects of dark matter been included the value of the baryonic, transition mass would have been reduced somewhat, but the corresponding dark matter mass would have approximated 10$^{12.5}$  M$_{\odot}$. In fact, the mass function of galaxies in the standard Press-Schechter parameterization declines exponentially above a certain critical mass, the stellar mass being roughly 10$^{11}$ M$_{\odot}$ and the corresponding halo mass being roughly 10$^{12.5}$  M$_{\odot}$. Consequently, we have both an observational and a physical basis for expecting that galaxies above some critical mass will grow primarily by accreting satellites and cannot be formed easily by a monolithic collapse. Thus, while matching based on $v_{max}$ will be best for normal systems, we can expect that, for first brightest galaxies in massive clusters, M$_{peak}$ should be the best metric.

\subsection{Rank Ordering in IllustrisTNG}\label{sec:ranking}

Here, we test these theoretical predictions using the IllustrisTNG simulation.  As described in Section \ref{sec:galaxy_selection}, we use a sample of galaxies with dark matter mass greater than 10$^{11}$ M$_{\odot}$/$h$ in the DMO simulation and stellar masses above 10$^9$ M$_{\odot}$/$h$ in the full hydrodynamical simulation. 

\begin{figure}
\begin{center}
\includegraphics[scale=0.53,trim=5mm 1mm 5mm 6mm, clip]{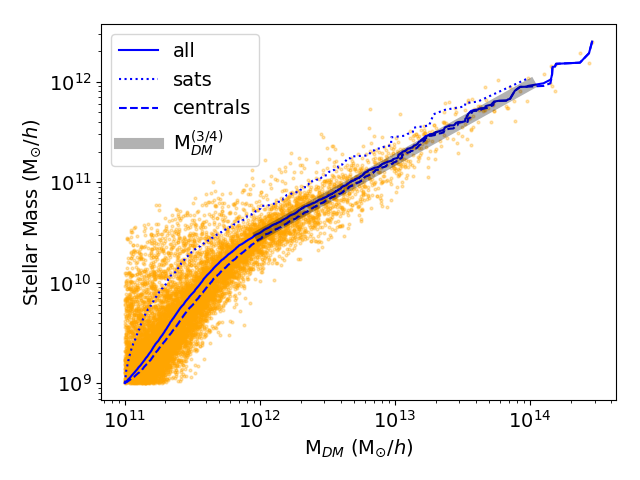}\\%{DM_SM_subhalos_selected_sorted_ulim_satsfix_veldisp_conc_illustris_TNGmatch_mstar9_dm11_dm9env_dmsat_mdm.png}\\
\includegraphics[scale=0.53,trim=5mm 1mm 5mm 6mm, clip]{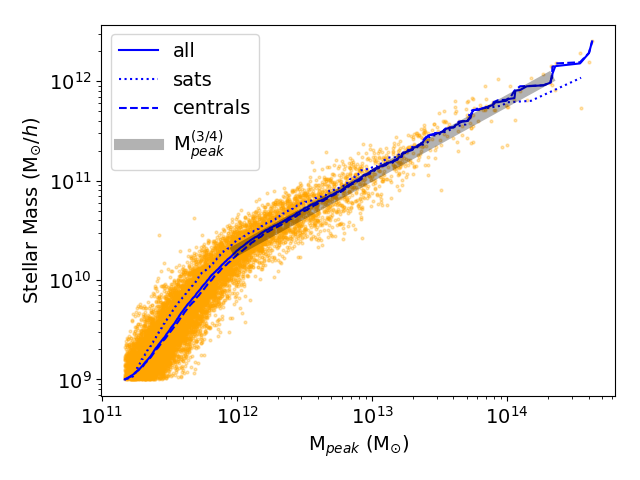}\\%{DM_SM_subhalos_selected_sorted_ulim_satsfix_vmax_M1Mpc_TNGmatch_mstar9_dm9env_dmsat_mpeak.png}\\
\includegraphics[scale=0.53,trim=5mm 4mm 5mm 6mm, clip]{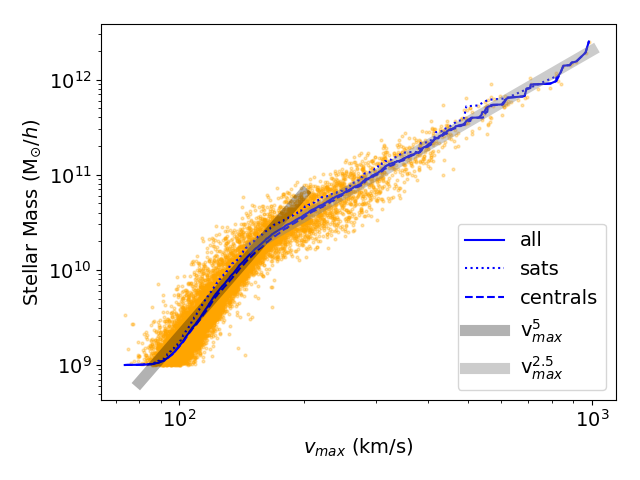}\\
\caption{The stellar mass of galaxies versus possible variables to use for ranking.  Blue lines show the predicted stellar mass using the rank-ordering method for each of our samples. \textbf{Top panel:}  Ranking using the current dark matter halo mass has the most scatter. \textbf{Middle panel:} Using M$_{peak}$ for ranking reduces scatter, and the rank ordering predictions for satellite and central galaxies is much closer.  \textbf{Bottom panel:} Ranking using $v_{max}$ reduces scatter dramatically, particularly for lower mass (lower $v_{max}$) halos, as quantified in Table 1. }\label{fig-dmsm}
\end{center}
\end{figure}

We rank-ordered our selected galaxies by total mass and the stellar mass separately for each of our samples:  "all" (11927), "satellites" (2337), and "centrals" (9590).  We have used three simple proxies for dark matter halo mass in our ranking schemes: the current dark matter mass, M$_{DM}$, the peak dark matter mass, M$_{peak}$, and the current $v_{max}$.  These are shown in order from the top to bottom panels in Figure \ref{fig-dmsm}.  We show the total mass proxy and stellar mass for each of our galaxies as orange ``o".  The blue lines show the predicted stellar mass using the rank-ordering method for each of our samples.  

We highlight that M$_*$ is the current stellar mass at $z$ $=$ 0, including any mass loss from star particles due to stellar evolution \citep{Vogelsberger2013MNRAS.436.3031V, Leitner2011ApJ...734...48L,Wiersma2009MNRAS.399..574W}.  Within $\sim$3 Gyr of formation, 45\% of a star particle's mass can be lost due to stellar evolution \citep{Leitner2011ApJ...734...48L}, inserting a factor of $\sim$2 into Equation \ref{eqn:mstarvmax4}.  However, we cannot simply include a constant term into the relation between current stellar mass and halo mass (or $v_{max}$) because of the time-dependence of stellar evolution, and the fact that mass recycled from earlier stellar populations may be required to form later generations of stars.  The time-dependence of stellar mass loss would move galaxies with earlier formation times to lower M$_*$ as they have had more time to recycle stellar mass back to gas mass.  However, we note that earlier formation times are correlated with higher stellar mass at a constant halo mass (\citealp[e.g.][]{Matthee2017}), so stellar evolution likely only flattens this relation.

We see that the scatter decreases as we move from using M$_{DM}$ to M$_{peak}$, although the M$_*$ $\propto$ M$_{halo}^{(3/4)}$ for higher masses holds for both variables.  We also highlight that the rank-order line for satellite galaxies is much closer to that for centrals when we use M$_{peak}$ than when we use M$_{DM}$.  The scatter continues to decrease when we use $v_{max}$ as our dark matter halo mass proxy, particularly at lower masses (lower $v_{max}$).  To guide the eye we have overplotted simple power-law relations between M$_*$ and $v_{max}$.

%fig 1

The results shown in Figure \ref{fig-dmsm} are quantified using the percent error as described in Section \ref{sec:error} (eqn 4), with the results shown in Table 1.   We see that while using the current M$_{DM}$ in the matching scheme is reasonably accurate for central galaxies, it is much less accurate for satellite systems.  Therefore, we also consider the peak mass of the halo, M$_{peak}$.  Using M$_{peak}$ should correct for mass loss from satellite galaxies due to tidal stripping.  Because dark matter is distributed to a larger radius than the stars in a galaxy, it will be more strongly stripped.  

Therefore, while we expect M$_{peak}$ to be very similar to M$_{DM}$ for central galaxies, it can vary by a considerable amount for satellites.  Indeed, we see in Table 1 that the improvement for centrals is very small when using M$_{peak}$ rather than M$_{DM}$, but it is dramatic for satellite galaxies.  We also consider $v_{max}$, as tidal stripping is found to have little effect on this property, likely because the maximum rotational velocity is reached at relatively low radii.  Using $v_{max}$ for our variable we find that the error for central galaxies has improved by more than 15\%, although the correlation between $v_{max}$ and stellar mass for satellites is somewhat weaker than the correlation between M$_{peak}$ and stellar mass.  However, because most of our galaxies are centrals, $v_{max}$ remains the best single variable for rank-ordering our galaxy sample.

We stress that, because of the shape of the mass function, any relation between stellar mass and halo mass will be dominated by the lowest-mass galaxies.  Therefore, we also consider separately only galaxies whose mass in the DMO simulation is greater than 10$^{12}$ M$_{\odot}/h$ in order to remove the bulk of low mass galaxies while still retaining a sample with $\sim$200 satellite galaxies.  

Unsurprisingly, we find that, as in the full sample, ordering using M$_{DM}$ is reasonably accurate for central galaxies, but much less so for satellite systems.  Again, we find a large improvement in the ranking scheme for satellite galaxies using M$_{peak}$.  

However, unlike in the full sample, $v_{max}$ is the worst ranking variable for central galaxies with halo masses above 10$^{12}$ M$_{\odot}/h$.  This agrees well with our theoretical argument that at large masses M$_{peak}$ will be the best ranking variable due to merging.

With this empirical support for the trends predicted in our model, we also develop a straightforward variable, using the physical intuition from above, that $v_{max}$ will be the best ranking variable for low mass galaxies and M$_{peak}$ will be the best ranking variable for high mass galaxies (Section \ref{sec:theory}).  For this variable we normalize both $v_{max}$ and M$_{peak}$ to their values at a ``pivot mass" of M$_{peak}$ $=$ 10$^{12.7}$ M$_{\odot}$.  We call these variables $v_{norm}$ $\equiv$ $v_{max}/v_{max, 12.7}$ and $m_{norm} \equiv M_{peak}/10^{12.7}$.  We then rank order our galaxies using the parameter based on these normalized values: 

\begin{equation}
\phi\equiv v_{norm} +  m_{norm} 
\end{equation}

Using this parameter, low mass galaxies depend more strongly on $v_{max}$, while high mass galaxies depend on M$_{peak}$. Both the exact value of the pivot mass and the powers of $v_{norm}$ and $m_{norm}$ were selected to minimize error while fleshing out our theoretical scaffold.  
%we also test our derived parameter, $\phi$$\equiv$$v_{norm}$ $+$  $m_{norm}$.  %, testing several $\alpha$ and $\beta$ values.  

As shown in Table 1, using this parameter $\phi$ gives some improvement on the fit to the central galaxies in our sample, and dramatically reduces the error for the satellite galaxies.  Using this variable for the mix of all galaxies reduces the error by a substantial 33\% when compared with rank ordering by M$_{DM}$. %, resulting in overall improvement of the matching scheme.  

\begin{table*}
\begin{center}
\label{tbl-RMSE}
\begin{tabular}{ c|c|c|c|c|c } 
 Number of galaxies (M$_{DM}$ $>$ 10$^{11}$ M$_{\odot}$) & 11927 & 9590 & 2337 & 11927 & 11927 \\
 \hline
Galaxy Sample & All & Centrals & Satellites & Mix & \% Improvement\\
 \hline
 Rank Ordering using M$_{DM}$ & 0.198 & 0.130 & 0.279 & 0.159 & --\\ 
 Rank Ordering using M$_{peak}$ & 0.136 & 0.127 & 0.133 & 0.128 & 19 \\ 
 Rank Ordering using $v_{max}$ & 0.116 & 0.106 & 0.137 & 0.112 & 30 \\ 
 Rank Ordering using $\phi$ $\equiv$ $v_{norm}$ + $m_{norm}$ & 0.111 & 0.101 & 0.119 & 0.105 & 34 \\
 %Rank Ordering using $v_{norm}^{2}$ + $m_{norm}^1$ & 0.112 & 0.102 & 0.118 & 0.106 \\
 %Rank Ordering using $v_{norm}^{3}$ + $m_{norm}^1$ & 0.113 & 0.104 & 0.118 & 0.107 \\
 %Rank Ordering using $v_{norm}^{4}$ + $m_{norm}^1$ & 0.115 & 0.107 & 0.117 & 0.109 \\
 %Rank Ordering using $v_{norm}^{1}$ + $m_{norm}^2$ & 0.112 & 0.102 & 0.125 & 0.106 \\
 %Rank Ordering using $v_{norm}^{1}$ + $m_{norm}^3$ & 0.114 & 0.103 & 0.132 & 0.109 \\
 %Rank Ordering using $v_{norm}^{2}$ + $m_{norm}^2$ & 0.112 & 0.102 & 0.126 & 0.106 \\
 %Rank Ordering using $v_{norm}^{2}$ + $m_{norm}^3$ & 0.114 & 0.103 & 0.133 & 0.109 \\
 %Rank Ordering using $v_{norm}^{0.5}$ + $m_{norm}^1$ & 0.112 & 0.103 & 0.118 & 0.106 \\
 %Rank Ordering using $v_{norm}^{1}$ + $m_{norm}^{0.5} $ & 0.114 & 0.104 & 0.117 & 0.107 \\
 %Rank Ordering using $v_{norm}^{0.5}$ + $m_{norm}^{0.5} $ & 0.116 & 0.106 & 0.117 & 0.108  \\
 %Rank Ordering using $v_{norm}^{1}$ + ($M_{DM}/M$_{pivot}$)^{1} $ & 0.129 & 0.102 & 0.165 & 0.115 \\
 
\end{tabular}
\\
Table 1: The percent error (eqn 4) of the ranking method using a single variable chosen to be either M$_{DM}$ (at $z=0$), M$_{peak}$, $v_{max}$ (at $z=0$), and $\phi$ $\equiv$ $v_{norm}$ + $m_{norm}$ (eqn 11) for the dark matter mass for the different samples (a fit to all galaxies, only centrals, only satellites, and mix of all galaxies fitting the centrals and satellites separately).  We use a galaxy sample with dark matter mass in the DMO simulation greater than 10$^{11}$M$_{\odot}/h$ that is matched to any galaxy in the hydrodynamical run with stellar mass greater than 10$^{9}$M$_{\odot}/h$.  The final column shows the percent improvement of ranking by the selected variable compared to M$_{DM}$ (at $z=0$) on the ``mix" sample.  We see that ranking using the single variable $\phi$ reduces the error 34\% compared to matching by M$_{DM}$ (at $z=0$).  %The number of nearby galaxies is actually nearby galaxies + 1, and the mass is mass of nearby halos + mass of dm halo.  The satellite versus central determination is made using the DMO simulation.
\end{center}
\end{table*}

%\begin{table*}
%\begin{center}
%\label{tbl-RMSE}
%\begin{tabular}{ c|c|c|c|c } 
% Number of galaxies (M$_{DM}$ $>$ 10$^{12}$ M$_{\odot}$) & 1659 & 1463 & 196 & 1659 \\
% \hline
%Galaxy Sample & All & Centrals & Satellites & Mix\\
% \hline
% Rank Ordering using M$_{DM}$ & 0.132 & 0.118 & 0.181 & 0.126 \\ 
% Rank Ordering using M$_{peak}$ &  0.120 & 0.118 & 0.122 & 0.119 \\
% Rank Ordering using M$_{peak}$ with 50 removed & 0.120 & 0.119 & 0.124 & 0.119 \\
% Rank Ordering using vmax & 0.124 & 0.123 & 0.129 & 0.124 \\ 
% Rank Ordering using vmax with 50 removed & 0.127 & 0.126 & 0.134 & 0.126 \\  
%\end{tabular}
%\\[5]
%Table 2: The percent error of the ranking method using current M$_{DM}$, Mpeak, and vmax for the dark matter mass for the different samples (a fit to all galaxies, only centrals, only satellites, and all galaxies fitting the centrals and satellites separately).  Using the galaxy sample with dark matter mass greater than 1$\times$10$^{12}$M$_{\odot}$/h that is matched to any galaxy in the hydro run with stellar mass greater than 1$\times$10$^{9}$M$_{\odot}$/h.  %The number of nearby galaxies is actually nearby galaxies + 1, and the mass is mass of nearby halos + mass of dm halo.  The satellite versus central determination is made using the DMO simulation.
%\end{center}
%\end{table*}

\begin{table*}
\begin{center}
\label{tbl-RMSE}
\begin{tabular}{ c|c|c|c|c|c } 
Number of galaxies (M$_{DM}$ $>$ 10$^{11}$ M$_{\odot}$) & 11927 & 9590 & 2337 & 11927 & 11927 \\
 \hline
 Galaxy Sample & All & Centrals & Satellites & Mix & \% Improvement\\
 \hline
 %Rank Ordering using M$_{DM}$ & 0.198 & 0.130 & 0.279 & 0.159 \\ 
 %Rank Ordering using M$_{peak}$ & 0.136 & 0.127 & 0.133 & 0.128 \\ 
 %Rank Ordering using vmax & 0.116 & 0.106 & 0.137 & 0.112 \\ 
% Rank Ordering using Concentration vratio (Jerry) & 0.741 & 0.717 & 0.687 & 0.711 \\
 %RF using only Concentration vmax & 0.462
 %RF using only Concentration vratio & 0.448
 %RF using only Concentration rratio & 0.450
 %RF using only veldisp & 0.122
 %RF using only M1Mpc &  0.320
 %RF using only M2Mpc &  0.403
 %RF using only vmax rad & 0.327
 %RF using only dm rad & 0.283
 %RF using only M8Mpc & 0.460
 
 %\hline
 \hline

 %vmax + veldispdm (f2) & 0.115 & 0.105 & 0.138 & 0.111 \\
 $\phi$ + $v_{disp}$ & 0.112 & 0.102 & 0.117 & 0.105 & 0 \\
 $\phi$ + $v_{max}$ & 0.111 & 0.101 & 0.117 & 0.104 & 1 \\
 %vmax + massdm (f2) & 0.114 & 0.102 & 0.136 & 0.109 \\
 $\phi$ + M$_{DM}$ & 0.105 & 0.101 & 0.110 & 0.103 & 2\\
 $\phi$ + M$_{peak}$ & 0.111 & 0.101 & 0.117 & 0.104 & 1\\
 %vmax + Mpeak (f2) & 0.111 & 0.101 & 0.121 & 0.105 \\
 \hline
 %vmax + vmax rad (f2) & 0.116 & 0.103 & 0.138 & 0.110 \\
 $\phi$ + r$_{max}$ & 0.111 & 0.101 & 0.118 & 0.105 & 0 \\
 %vmax + dm rad (half mass) (f2) & 0.114 & 0.103 & 0.135 & 0.109 \\ %halfmass radius
 $\phi$ + r$_{DM}$ & 0.105 & 0.100 & 0.114 & 0.103 & 2 \\ %halfmass radius
 \hline
 %vmax + Concentration vmax (Illustris paper) (f2) & 0.116 & 0.103 & 0.138 & 0.110 \\
 $\phi$ + $c_v$ & 0.111 & 0.101 & 0.118 & 0.105 & 0 \\
 %vmax + Concentration vratio (Jerry) (f2) & 0.112 & 0.104 & 0.135 & 0.110 \\
 $\phi$ + $c_h$ & 0.109 & 0.101 & 0.117 & 0.104 & 1 \\
 $\phi$ + $c_r$ & 0.109 & 0.101 & 0.116 & 0.104 & 1\\
 %vmax + Concentration rratio (R$_{vmax}$/R$_{halfmass}$) (f2) & 0.113 & 0.105 & 0.133 & 0.111\\
 \hline
 $\phi$ + $t_{peak}$ & 0.105 & 0.101 & 0.110 & 0.103 & 2 \\
 $\phi$ + $t_{50}$ & 0.106 & 0.099 & 0.116 & 0.102 & 3\\
 $\phi$ + $t_{85}$ & 0.104 & 0.099 & 0.111 & 0.101 & 4 \\
 \hline
 %vmax + time since Mpeak (f2) & 0.113 & 0.106 & 0.130 & 0.111 \\
 %vmax + t50 & 0.116 & 0.106 & 0.137 & 0.112 \\
 %vmax + t85 & 0.114 & 0.106 & 0.131 & 0.111 \\
%\hline

 $\phi$ + M$_{DM, r < 1 Mpc}$ & 0.104 & 0.100 & 0.115 & 0.103 & 2 \\

% vmax + DM Env (M$_{DM}$$>$10$^{9}$)number in 1 Mpc (f2) & 0.107 & 0.101 & 0.129 & 0.107 \\
%\hline
 %vmax + DM Env (M$_{DM}$$>$10$^{9}$) mass in 2 Mpc (f2) & 0.110 & 0.103 & 0.132 & 0.109 \\
 $\phi$ + M$_{DM, r < 2 Mpc}$ & 0.103 & 0.099 & 0.113 & 0.102 & 3 \\
 %vmax + DM Env (M$_{DM}$$>$10$^{9}$) number in 2 Mpc (f2) & 0.109 & 0.103 & 0.132 & 0.108 \\
 %\hline
 %vmax + DM Env (M$_{DM}$$>$10$^{9}$) mass in 5 Mpc (f2) & 0.112 & 0.104 & 0.134 & 0.110 \\
 $\phi$ + M$_{DM, r < 5 Mpc}$ & 0.105 & 0.099 & 0.115 & 0.102 & 3\\
 %vmax + DM Env (M$_{DM}$$>$10$^{9}$) number in 5 Mpc (f2) & 0.112 & 0.104 & 0.134 & 0.110\\
 %\hline
 %vmax + DM Env (M$_{DM}$$>$10$^{9}$) mass in 8 Mpc (f2) & 0.114 & 0.104 & 0.135 & 0.111 \\
 $\phi$ + M$_{DM, r < 8 Mpc}$ & 0.107 & 0.100 & 0.116 & 0.103 & 2\\
% vmax + DM Env (M$_{DM}$$>$10$^{9}$) number in 8 Mpc (f2) & 0.114 & 0.105 & 0.135 & 0.111\\
% \hline
 %vmax + DM Env (M$_{DM}$$>$10$^{9}$) mass in 15 Mpc (f2) & 0.115 & 0.105 & 0.136 & 0.111 \\
 $\phi$ + M$_{DM, r < 15 Mpc}$ & 0.109 & 0.100 & 0.117 & 0.104 & 1\\
 %vmax + DM Env (M$_{DM}$$>$10$^{9}$) number in 15 Mpc (f2) & 0.115 & 0.105 & 0.136 & 0.111\\
 \hline
 \hline
 $\phi$ + M$_{DM, r < 2 Mpc}$ + $t_{85}$ & 0.101 & 0.097 & 0.108 & 0.099 & 6 \\
 %vmax + DM Env (M$_{DM}$$>$10$^{9}$) mass in 1 Mpc + Mpeak & 0.105 & 0.100 & 0.116 & 0.103\\
 %vmax + DM Env (M$_{DM}$$>$10$^{9}$) mass in 8 Mpc + Mpeak & 0.107 & 0.099 & 0.117 & 0.103\\
 %vmax + DM Env (M$_{DM}$$>$10$^{9}$) mass in 1 Mpc + Concentration vratio & 0.107 & 0.100 & 0.130 & 0.106\\
 \end{tabular}
\\
Table 2: The percent error using two variables: the $\phi$ ranking method plus the listed corrections.  We use the galaxy sample with dark matter mass greater than 10$^{11}$M$_{\odot}/h$ that is matched to any galaxy in the hydro run with stellar mass greater than 10$^{9}$M$_{\odot}/h$.  Note that M$_{DM, r < X Mpc}$ is the mass of all halos within that radius from the DMO simulation with M$_{DM}$ $>$ 10$^{9}$M$_{\odot}$, including the mass of the halo from which the measurement originates.  All halo properties are measured using the DMO simulation.  Here the final column shows the percent improvement on the ``mix" sample of using the correction variable in addition to rank-ordering by $\phi$ in comparison to only rank-ordering by $\phi$.
\end{center}
\end{table*}

\section{Using Secondary Variables to Improve Rank Ordering}\label{sec:secondary_variables}

We now attempt to minimize the scatter in the $\phi$ - M$_*$ relation using other features of dark matter halos.  These features are listed in Table 2.  We have roughly grouped the halo properties into those related to the halo mass (M$_{DM}$, M$_{peak}$, $v_{disp}$$\equiv$dark matter velocity dispersion, and $v_{max}$), size (r$_{max}$$\equiv$ $v_{max}$ radius and r$_{DM}$$\equiv$ dark matter half mass radius), shape (concentration using the three methods described in Section \ref{sec:concentration}), formation time (the lookback time to M$_{peak}$, to when the halo reaches 50\% of its z=0 mass, and to when the halo reaches 85\% of its z=0 mass), and surrounding environment (using the mass of dark matter halos within various radii). 

In this section we first describe the method we use to include secondary features, and then discuss the results.  %highlight the results of two of the most useful features when rank ordering by $v_{max}$:  M$_{peak}$ and the mass in halos within 1 Mpc.  

\subsection{Method of Correction}\label{sec:correction_method}

\begin{figure}
    \centering
    \includegraphics[scale=0.55]{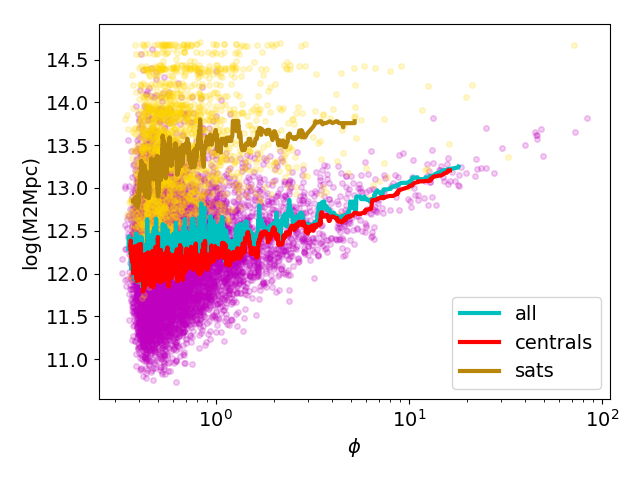}\\
    \includegraphics[scale=0.55]{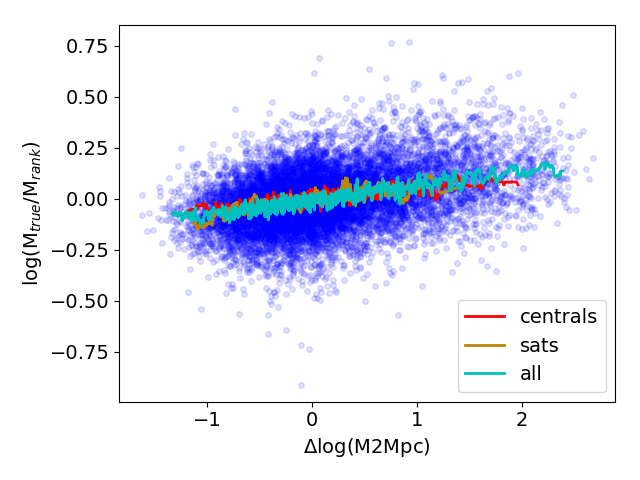}
    \caption{The top and bottom panels show the first and second steps used to include a secondary halo feature to reduce the scatter in rank ordering halos (Section \ref{sec:correction_method}).  Here we use the M$_{DM, r < 2 Mpc}$ environmental measure, written as M2Mpc.  \textbf{Top:} First we plot this variable as a function of $\phi$, our rank-ordering variable.  Here we show the total sample (``all") as well as the centrals and satellites ("sats") separately.  The points are color-coded as centrals and satellites.  \textbf{Bottom:} Using the scatter from a rolling median, we can find that the ratio of the true stellar mass of the galaxy to the rank-ordered assigned mass has a dependence on M$_{DM, r < 2 Mpc}$.  The points are not color-coded for satellites and centrals, as for the ``all" fit we use all of the galaxies in the sample.  We can then correct our stellar mass using this dependence.  Notice that the rolling median for the total sample is similar to that for the separated satellite and central samples.}
    \label{fig:correction}
\end{figure}

We first plot our feature as a function of our best single variable $\phi$, and find the running median of the feature using a window size of 50 galaxies.  We have tested using other window sizes (25 and 100 galaxies as well as a constant $\Delta log(\phi)$ bin of $\pm$0.3 around each galaxy) and find similar results in our percent improvement.  The top panel of Figure \ref{fig:correction} shows this plot using the environmental density M$_{DM, r < 2 Mpc}$ variable.  Clearly there is a trend of increasing environmental density as a function of $\phi$, and it differs for satellites and centrals.

Because we use the rolling median, we need to remove the first and last 25 values, so we are left with an "all" sample of 11877, a "centrals" sample of 9540, and a "satellites" sample of 2287 galaxies.  Removing these galaxies has little impact on the percent errors using the rank ordering method for each sample (a change of less than 1\%).  

We then plot M$_{true}$/M$_{rank}$ as a function of $\Delta$log(\textit{feature}), which is the difference between the log(\textit{feature}) for each dark matter halo and the log(\textit{feature}$_{rolling median}$) found at each $\phi$.  In the bottom panel of Figure \ref{fig:correction} we show how M$_{true}$/M$_{rank}$ is related to the scatter in M$_{DM, r < 2 Mpc}$.  This relation is fit using first, second and third order polynomial fits.  Finally, we correct our prediction for the stellar mass using our chosen fit as below (the second order polynomial fit shown tends to give the best results):

\begin{multline}
log(M_{*,pred}) = log(M_{*,rank}) + \\
\alpha\Delta log(feature)^2 + \beta \Delta log(feature) + \gamma
\end{multline}

Finally, we calculate the percent error of the new prediction.  This value for each feature and galaxy population is shown in Table 2.

\subsection{Results}

All of our quantitative results are shown in Table 2.  The most glaring result is that most corrections do not result in a large improvement of the percent error from ranking using $\phi$.  

Using random resampling of 70\% of our data sets (``all",``central", and ``satellites") 60 times, we find a distribution of errors with means matching the values listed for $\phi$ of the complete sample in Table 1, and standard deviations of 0.001, 0.001, 0.002, and 0.0009 for ``all", ``centrals", ``satellites" and ``mix" samples, respectively.  With this in mind we can look more closely at the improvement when adding a second feature to our matching scheme.

In some more detail, it is not surprising that all of the halo features describing halo mass do not improve the fit to central galaxies at all.  These are well-fit by our $\phi$ variable.  However, interestingly, the error is reduced for the satellite sample when we include a M$_{DM}$ correction.  This may be because we have largely ignored satellite galaxy evolution by choosing $v_{max}$ and M$_{peak}$ as the components of $\phi$.  Including M$_{DM}$ may start to include the later evolution of these galaxies.  

We find universally small improvement when considering our variables describing halo size (r$_{max}$ and r$_{DM}$) and shape (concentration).  Although previous work has shown that the scatter in abundance matching is related to concentration (\citealp[e.g.][]{Matthee2017}), it is not surprising that concentration does not improve the scatter when ranking using our $\phi$ variable.   This is because $\phi$ includes $v_{max}$, which is already related to concentration \citep{Klypin2011}.

Interestingly, there is some improvement in the error when folding formation time into the stellar mass estimate.  For example, $t_{85}$ is the halo feature that results in the smallest percent errors across all of our samples: ``all" galaxies, ``centrals", ``satellites", and the ``mix" sample.  As discussed above, although we find that earlier formation times are correlated with higher stellar mass at a constant $\phi$ (in agreement with previous work), stellar evolution causing mass loss over several Gyr may flatten this relationship.

Finally, using environment to correct for the stellar mass also has a small impact on the overall error.  Despite this, we note that including the mass from galaxies within 2-5 Mpc seems to produce a slightly better correction than smaller or larger environment windows.

\subsubsection{Correcting Using A Combination of Environment and Formation Time}

Finally, we use our fits for each of our strongest individual corrections, $t_{85}$ and M$_{DM, r < 2 Mpc}$, to create a combined correction on the rank-ordering technique.

\begin{equation}
\begin{array}{l}
M_{*,pred} = log(M_{*,rank}) +  \\
(\alpha_{M2}\Delta log(M_{DM, r < 2 Mpc})^2 + \beta_{M2} \Delta log(M_{DM, r < 2 Mpc}) +\\
 (\alpha_{t85}\Delta log(t_{85})^2 + \beta_{t85} \Delta log(t_{85}) + \gamma
\end{array}
\end{equation}

We use the curvefit module in scipy to perform a least-squares fit to the above equation, and find that we can reduce the error using both M$_{DM, r < 2 Mpc}$ and $t_{85}$ as shown in the final line of Table 2.

\begin{table*}
\begin{center}
\label{tbl-RMSE}
\begin{tabular}{ c|c|c|c|c } 
 & All & Centrals & Satellites & Mix\\
 \hline
 $\phi$ & 0.111,0.111 & 0.102,0.101 & 0.120,0.117 & 0.105,0.104\\
 $\phi$ + M$_{DM, r < 2 Mpc}$ & 0.104,0.104 & 0.099,0.099 & 0.114,0.112 & 0.102,0.101\\
 $\phi$ + $t_{85}$ & 0.104,0.104 & 0.099,0.100 & 0.111,0.109 & 0.101,0.102\\
 $\phi$ + M$_{DM, r < 2 Mpc}$ + $t_{85}$ & 0.101,0.100 & 0.097,0.097 & 0.109,0.104 & 0.099,0.097\\
  \end{tabular}
\\
Table 3: The median percent error on ten iterations using a training and test sample. For each sample and fitting method we list the training then test sample values.
\end{center}
\end{table*}
%\subsection{Improving the High Mass Sample}

\subsection{Verifying our Results}

Here we use two methods to verify our results on the improvement using multiple halo features to determine stellar mass. 

\subsubsection{Random Forest Regression}

Now that we have gained insight into the level of improvement that can be gained by using more than one feature of dark matter halos in the abundance matching technique, we turn to machine learning to provide an independent check of our modeling and ranking scheme.  

Using Random Forest Regression (RFR) allows us to rank the features according to their effect on the model output, and has the additional benefit of expanding the space of available models beyond polynomial fitting. %allows us to find the likely optimal number of features to reduce our error as much as possible.  Further, we are not limited to fitting curves to functions.  
For this work we use scikit-learn \citep{Pedregosa2011}.

First, we are able to reproduce the percent error on the entire sample using only our defined $\phi$ feature (0.111), and  in a two-feature setting where we add a central/satellite galaxy label (0.105).  We check the rest of our ranking parameters from Table 1 and verify that $\phi$ produces the best ranking variable to match DM halos to galaxies.  Also, we confirm that our $\phi$ variable produces lower error values than the combination of $v_{max}$ and M$_{peak}$.  %vmax gives 0.116, Mpeak gives 0.135, and MDM gives 0.188

We also use our four selected halo features that we found produced the best match between the DMO and hydrodynamical simulations, $\phi$, M$_{DM, r < 2 Mpc}$, $t_{85}$, and the central/satellite label.  Using an optimized RF regressor, the expected test set error is 0.098 with a standard deviation of 0.0013, quite similar to the percent error we find ranking the satellites and centrals separately using $\phi$ and applying our analytic correction using M$_{DM, r < 2 Mpc}$ and $t_{85}$. This is reassuring because it shows that our results are only very mildly dependent on the modeling assumptions.

We can also include all the features and use a parameter optimization technique to find the minimum possible error of a Random Forest Regression.  We find a minimum error of 0.092 using eight randomly selected features, creating 100 trees (n$_{estimators}$) with a maximum depth of 14 branches (max$_{depth}$).  However, there are more than 30 combinations within one standard deviation (0.0015), including one using only 4 features.  We can conclude that there may be many similarly relevant predictors in our feature list.  This supports our analytic reasoning that several of our halo features are reasonable proxies for halo mass, and we have already noted that our other halo features can be separated into only a few types of variables (halo size, concentration, formation time, and environment).

Indeed, if we optimize the RFR including one feature of each type we can reach an error of 0.095 ($\phi$, M$_{DM, r < 2 Mpc}$, $t_{50}$, r$_{DM}$, and $c_v$).  This is within two standard deviations of the four halo parameters we use in our analytic model, and so does not indicate a dramatic improvement.

Comparing our results to the errors found using the RFR machine learning technique gives us assurance that our analytic method for including extra halo features is reasonable, and that our conclusions are not strongly model-dependent.  While continuing to add features can reduce the error on the matching scheme, we do not find other clear DM halo features that dramatically improve upon our analytic method.

\subsubsection{Cross-Validation}

In order to obtain another view on whether increasing the number of halo features improves our estimate of stellar mass we can use cross-validation.  This can be used to determine how meaningful our derived improvements are when used to \textit{predict} the stellar mass of galaxies.  Cross-validation is specifically designed to trade off over- and under-fitting to give the highest prediction accuracy.  For this, we randomly select 80\% of our sample as our test set, on which we perform the fitting processes as described.  We use the remaining 20\% as our test set to determine if the percent error on the stellar mass prediction improves when including more features.  Specifically, we select 80\% of our total sample for the ``all" fits, and then 80\% of the central and satellite samples, in order to determine the ``central", ``satellite", and ``mix" fits.  

We performed this cross-validation routine ten times using ten different random subsets of the data, and universally find improvement in both the training and test sets when using M$_{DM, r < 2 Mpc}$, $t_{85}$, or their combination.  In Table 3, we list the median percent error values for the ten sets of training and test samples.  We can conclude that we have not yet overfit using these halo features, and our improvement in predicting stellar masses from halo masses is real, albeit small. 

\section{Discussion}\label{sec:discussion}

What have we learned from this exercise? The zeroth order conclusion is that a matching scheme based on the maximum velocity in a dark matter halo is a good single predictor of the final stellar mass for normal galaxies, whether they are central galaxies or satellites. The typical error in the prediction (in the IllustrisTNG100 simulations) is 11.6 percent in log(M$_*$) compared to 19.8 using M$_{DM}$, and the dependence of stellar mass on $v_{max}$ is unsurprisingly log (M$_*$) $\sim$ (3.8 $\pm$ 0.02) log($v_{max}$) (using bootstrap resampling with 70\% of the data set), close to the Faber-Jackson relation \citep{Faber1976}. This result is just what one would have expected from the simplest physical argument that estimates the amount of gas that can be turned into stars in the standard \citet{Gunn1972} collapse of a dark matter halo. 

But, for high mass systems comparable to the first brightest galaxies in clusters living in halos more massive than 10$^{12.7}$ M$_{\odot}$, the accretion of satellite systems will significantly increase the stellar mass and the most relevant halo parameter is simply the peak dark matter mass, M$_{peak}$.  Using a single variable, $\phi$ (Equation (11)), which incorporates both features reduces the error to 10.5\% when satellites and centrals are ranked separately.  

These prescriptions should be easy to implement and can replace the simplest, halo mass based initial matching schemes when estimating the expected galaxy stellar masses given a dark matter simulation. 

If one wants to go farther and improve the best zeroth order scheme by first order corrections then we have found that a roughly 6\% improvement is possible.  Interestingly, environmental considerations that we considered did not lead to significant improvement even in satellite galaxies, and the best single variable for improvement was $t_{85}$, the time at which a halo reached 85\% of its peak mass.

However, an almost mindless combination of the two variables ($v_{max}$, M$_{peak}$) worked best. We further found that a simple linear combination based on these two variables enables predictions to a typical accuracy of 10.5 percent error in log(M$_*$).

\section{Tests}\label{sec:tests}

All of these results are based on simulated data and it is important to test them in the real world.  We have been able to think of two tests that might be applied to help determine whether the proposed matching scheme provides a significant improvement over the simplest matching scheme.  First one constructs a standard LCDM, dark matter only simulation and, using a standard halo finding algorithm, makes a catalog of dark matter halos labeling each of them with the final dark matter mass, M$_{DM}$, the peak dark matter mass M$_{peak}$ over the history of the halo and the current halo maximum circular velocity $v_{max}$. Then, to test the classic matching scheme (as has been done before – \citealp[][]{Conroy2006}), one takes a representative volume and rank orders the halos by M$_{DM}$, takes catalog values for a comparable volume (from, say, the Sloan Digital Sky Survey) and rank orders the observed galaxies by (for example) $g$ or $r$ magnitudes and then identifies each DM halo with the matched by ranking, real galaxy. This gives one an artificial catalog of galaxies each tagged with a position, a velocity and a $g$ or $r$ optical magnitude.

Then one would “observe” this synthesized catalog and construct two spatial distribution functions, a galaxy-galaxy spatial correlation function \citep{Conroy2006, Hearin2013a, Hearin2013b, Reddick2013} and a void distribution function (\citealp[e.g.][]{Walsh2019MNRAS.488..470W}). These could then be compared to the known galaxy-galaxy spatial correlation functions and the known void distribution functions with both one parameter functions specified as a function of magnitude. The magnitude distribution itself is of course correct by construction.
Then comparing – say – the autocorrelation length as a function of galaxy magnitude between the real and synthesized data sets allows one to determine the fractional error as a function of galaxy magnitude. %This would be the error of the standard matching algorithm, as has been studied by (X,Y,Z- REFs).% with typical errors of roughly XX % in the two spatial functions.

Then one would go back to the original DM halo catalog and, using (M$_{peak}$,$v_{max}$), construct for each halo the value of $\phi$ = ($v_{max}$/$v_{1}$) + (M$_{peak}$/(M$_1$)) (Equation 11), where ($v_1$, M$_1$) are the values of ($v_{max}$, M$_{peak}$) for the average halo of mass 10$^{12.7}$ M$_{\odot}$. Now, with each halo tagged with its value of $\phi$, one can rank order the synthetic sample by $\phi$ and attach visual magnitudes to each galaxy by the same method as was done using M$_{DM}$. Now one has a new catalog to observe with respect to spatial distribution metrics and can again find the fractional error in – for example – the spatial autocorrelation length as a function of visual brightness and compute the error by comparing to real observed data.

This procedure would give us a quantitative estimate as to how well the matching scheme was working compared to both reality and the previous simpler matching scheme which has had considerable success. And, unlike the exercises in this paper, the tests would not be dependent on the accuracy of our current galaxy formation algorithms, which, while well tested, suffer from the ``confirmation bias'' inevitable when uncertain modelling parameters are adjusted to fit observations.
We look forward to pursuing these independent tests in future work.

\section{Conclusions}\label{sec:conclusions}

In this paper we have examined schemes to populate a synthesized dark matter only set of cosmological simulations with galaxies to see if we could devise a simple and accurate scheme. We took as our starting point a matching scheme \citep{Vale2004,Vale2006,Vale2008} which, while almost naively simple has had some success. In that scheme, one rank orders DM halos by final mass and rank orders real galaxies in a similar cosmic volume by luminosity and attaches to the kth ranked halo the kth ranked galaxy. Table 1 represents of one parameter efforts which we compared to the computed luminosities in the IllustrisTNG simulated galaxy catalog.  Table 2 summarizes our results with two parameter fits where we used combinations of velocity dispersion, mass and environmental density. We did not find that adding an environmental variable produced a significant improvement over simpler schemes nor did we find that any of the two parameter fits that we investigated were statistically significantly superior to the one parameter fits. What we did discover was that a new single variable, $\phi$ (cf equation 12), which combines information from both mass and velocity variables, provides a quite significant improvement over the basic ranking scheme using final dark matter mass, the error being reduced by about 33\% percent. In our examination of the physical basis for the success of this new variable we examined simple arguments starting with the over half century old paper by \citet{Rees1977}.

There is a critical mass for galaxies – the mass above which it cannot cool by normal radiative processes in roughly a free fall time. That mass corresponds roughly to 10$^{12.7}$ M$_\odot$ which we designate as M1. Below this mass there is a simple analytic argument that asks if a gaseous object can cool in its own free fall time and is equivalent to M$_*$ $\sim$ $v_{max}^3f$. For masses above M1 growth only occurs by accretion of satellites and that is proportional to M. So, we designed a metric, $\phi$, which is dominated by velocity for low mass objects and dominated by mass for high mass objects more massive than M1. This single variable, based on the physical motivation given above, seems to provide a matching scheme superior to others which we have tested. We did try other combinations of (M$_{peak}$,$v_{max}$) and found none superior to the simple variable, $\phi$, that we had tested. So our bottom line is that the variable, $\phi$ (Equation 11), is the best single variable to use in predicting the stellar mass of galaxies, given their halo properties.

\acknowledgements
We would like to thank the referee for helpful comments that improved the paper.  ST would like to thank Claire Kopenhafer and Tjitske Starkenburg for their help and scripts in reading in and analyzing TNG outputs, Viviana Aquaviva for her Machine Learning class and comments on the draft, and Dan Foreman-Mackey for discussions and comments on cross-validation.
ST gratefully acknowledges support from the Center for Computational Astrophysics at the Flatiron Institute, which is supported by the Simons Foundation.  The data used in this work were hosted on facilities supported by the Scientific Computing Core at the Flatiron Institute, a division of the Simons Foundation. 

\clearpage 

\bibliography{TNGSHAM_bib}

\end{document}